\documentclass{siamltex}
\usepackage{epsfig}
\usepackage{amsmath}
\usepackage{amsfonts}
\def\bE{\mathbf E}

\def\bV{\mathbf V}

\def\bx{\mathbf x}
\def\by{\mathbf y}

\def\bk{\mathbf k}

\def\b0{\mathbf 0}
\def\bTheta{\mathbf \Theta}

\title{Singular Modes of the Electromagnetic Field\thanks{This research 
was financed by the Netherlands Organization for Scientific Research (NWO)
and by the Russian Foundation for Basic Research (RFBR).}}

\author{
Neil V. Budko\thanks{Laboratory of Electromagnetic Research, Faculty of Electrical Engineering, Mathematics 
and Computer Science, Delft University of Technology,
Mekelweg~4, 2628~CD, Delft, The Netherlands,
{\tt n.v.budko@tudelft.nl}}
\and Alexander B. Samokhin\thanks{Department of Applied Mathematics, Moscow Institute of
Radio Engineering, Electronics, and Automatics (MIREA), Verndasky~av.~78, 117454, Moscow, Russia.
{\tt absamokhin@yandex.ru}}
}

\begin{document}

\maketitle

\begin{abstract}
We show that the mode corresponding to the point of essential spectrum of the electromagnetic scattering 
operator is a vector-valued distribution representing the square root of the three-dimensional Dirac's delta function. 
An explicit expression for this singular mode in terms of the Weyl sequence is provided and analyzed. 
An essential resonance thus leads to a perfect localization (confinement) of the electromagnetic field, 
which in practice, however, may result in complete absorption. 
\end{abstract}



\pagestyle{myheadings}
\thispagestyle{plain}
\markboth{N.~V.~BUDKO AND A.~B.~SAMOKHIN}{SINGULAR MODES OF THE ELECTROMAGNETIC FIELD}

\section{Introduction}
%%%%%%%%%%%%%%%%%%%%%%%%%%%%%%%%%%%%%%%%%%%%%%%%%%%%%%%%%%%%%%%%%%%%%%%%%%%%%%
%%%%%%%%%%%%%%%%%%%%%%%%%%%%%%%%%%%%%%%%%%%%%%%%%%%%%%%%%%%%%%%%%%%%%%%%%%%%%%
The ability to manipulate the spatial distribution of the electromagnetic field is required in many practical applications. 
For example, one may wish to create an antenna with a very broad or a very narrow radiation pattern,
localize and amplify light or transmit it along a predefined optical path, accelerate charged particles or keep 
them tight within a fusion chamber. In the presence of matter all possible spatial distributions are encoded in the spatial spectrum
of the electromagnetic scattering operator. The most basic field distributions, which are easy to observe in 
microwave resonators, are called eigenmodes. Manipulation thus amounts to a clever excitation of a particular eigenmode
or a combination of those. If almost all electromagnetic energy is carried by one of the modes, then we talk about a resonance.
The concept of electromagnetic resonances and eigenmodes is a natural description of microwave resonators and waveguides \cite{Collin}, 
microstrip lines \cite{AmariTriki2003}, and other simple, often, infinite, homogeneous, 
or periodic structures \cite{KartchevskiNosichHanson2005}, \cite{FigotinKuchment1998}, \cite{ShipmanVenakides2003}. 
In a recent paper \cite{BudkoSamokhinPRL2006} we have generalized this idea for arbitrary dielectric objects of finite extent. 
The strongest point of our generalization was the incorporation of the full information about the spatial spectrum of the 
electromagnetic scattering operator \cite{BudkoSamokhinSIAM2006}, which has both discrete eigenvalues 
and an essential (continuous) part.

In some applications, e.g. optoelectronics, plasmonics, photonics, near-field optics, the electromagnetic field has to be confined 
within a very small volume of space, see e.g. \cite{MartinGirardDereux1995}, \cite{Maksimenko2000}, \cite{Akahane2003} . 
An observation made in \cite{BudkoSamokhinPRL2006} suggests that the modes associated
with the essential spectrum of the scattering operator may be highly localized in space. In particular, we argued that the
resonance, which corresponds to the essential spectrum, coincides with the so-called plasmon observed at a
plasma-dielectric interface. In research on metamaterials, where one strives for a negative permittivity material with 
vanishing losses, an unusual ``anomalous localized resonance'' is also encountered \cite{Milton2002}, \cite{Milton2005}.
Here we give a rigorous proof of the link between the localization or confinement of the electromagnetic field and
the essential spectrum of the electromagnetic scattering operator.

Our previous work \cite{BudkoSamokhinPRL2006}, \cite{BudkoSamokhinSIAM2006} was based on the Mikhlin's theory 
of singular integral operators \cite{Mikhlin2}, and does not concern with the shape of the modes. In fact, there are very 
few examples of the analysis of the modes corresponding to the essential spectrum of operators. In quantum mechanics, 
where the spectral theory is very advanced, the discrete spectrum seems to be of more importance. There the essential 
spectrum is associated with the unbounded motion of particles \cite{HislopSigal1996} and is, therefore, of little interest 
to physicists. In electromagnetics, however, it is the practical question of light confinement which is related to the essential 
spectrum, and the precise mathematical description of the associated modes is rather important. 

To recover the structure of the modes corresponding to the electromagnetic essential spectrum we resort here to the
Weyl's definition of spectrum, see e.g. \cite{HislopSigal1996}, \cite{DemuthKrishna2005}. This definition states that a number $\lambda$ is in the spectrum of operator $A$ if and only if there
exists a sequence $\{\Psi_{n}\}$ in the space $X$ such that 
 \begin{align}
 \label{eq:NormOne}
 \Vert \Psi_{n} \Vert = 1,
 \end{align}
and
 \begin{align}
 \label{eq:DefWeyl}
 \lim\limits_{n\rightarrow\infty} \Vert A\Psi_{n} - \lambda\Psi_{n}\Vert =  0.
 \end{align}
Furthermore, $\lambda$ is in the {\it essential spectrum}, if there is a {\it singular sequence} satisfying (\ref{eq:NormOne})--(\ref{eq:DefWeyl}), i.e. 
a sequence which contains no convergent subsequence. If $X$ is a complete Hilbert space, then, obviously, such singular sequence will not (strongly) converge to 
any function from $X$, although, it may weakly converge to zero. As we know, sequences that do (strongly) converge to some function on $X$ 
generate eigenfunctions or {\it eigenmodes} corresponding to the point spectrum -- {\it eigenvalues}. 
By analogy we may associate {\it essential modes} with the essential spectrum. An alternative term, which, perhaps, better reflects the
nature and structure of the particular modes obtained here, would be {\it singular modes}.

\section{The square root of the delta function}
%%%%%%%%%%%%%%%%%%%%%%%%%%%%%%%%%%%%%%%%%%%%%%%%%%%%%%%%%%%%%%%%%%%%%%%%%%%%%%
%%%%%%%%%%%%%%%%%%%%%%%%%%%%%%%%%%%%%%%%%%%%%%%%%%%%%%%%%%%%%%%%%%%%%%%%%%%%%%
Before going into the details of the electromagnetic case we shall introduce a peculiar function that is later used to generate the essential modes.
On one hand, the sequence of such functions should be singular, as required by the definition mentioned above. 
On the other hand, application of (\ref{eq:DefWeyl}) to electromagnetics (in $L_{2}$-norm) and consistency with the 
previously obtained results \cite{BudkoSamokhinSIAM2006} require that the square of this function should behave like the Dirac delta 
function, i.e., must have the sifting property. Hence, what we need is a {\it square root of the delta function}. 

From time to time the square roots of delta functions appear in literature. Mostly, though, just as a curious example of a non-convergent 
sequence, see e.g. \cite{Byron1992} (p.~299) and \cite{Peres1993} (p.~81). A more modern and rigorous approach to such functions 
is the Colombeau algebra \cite{Colombeau1992}, where one studies the products of distributions 
and encounters $m$-{\it singular delta functions}, which are almost identical to what we are after. At present, the main 
applications of the Colombeau algebra are: nonlinear equations (e.g.
hydrodynamics, elastodynamics and general relativity), singular shock waves in nonlinear conservation laws, and propagation of delta-like
waves in linear media with discontinuous parameters. In other words, this algebra is applied whenever a potentially meaningless product of 
generalized functions is stumbled upon. The present apparent emergence of the Colombeau algebra in the recovery of the essential 
spectrum seems to be new and can, probably, be generalized. This paper, however, does not attempt such a generalization, nor should it 
be considered an introduction to the Colombeau algebra. 

Unfortunately, we could not find any explicit derivation of the required distribution 
in the literature. The available one-dimensional and scalar three-dimensional \cite{HislopSigal1996} (pp.~74--75) examples are of no use to us, 
since the electromagnetic essential spectrum is a purely three-dimensional phenomenon and our function and its Fourier transform had to have 
a very special vectorial structure. In view of the potential usefulness of the obtained result in other areas of research, we have decided to 
devote this entire section to the analysis of the square root of the three-dimensional delta function. The proof of the following theorem is, however, 
rather technical and can be skipped in the first reading.

\begin{theorem} 
\label{th:PsiProperties} 
The vector-valued function 
\begin{align}
\label{eq:DefPsi}
\Psi(\alpha,\bx,\bx_{\rm c})=\left(\frac{2}{3}\right)^{1/2}\pi^{-3/4}\alpha^{5/4}(\bx-\bx_{\rm c}) 
\exp\left(-\frac{\alpha}{2}\vert\bx-\bx_{\rm c}\vert^{2}\right),
\end{align}
where $\bx,\bx_{\rm c}\in{\mathbb R}^{3}$ and $\alpha\ge 0$, has the following properties:
\begin{enumerate}
\item{It is normalized in the sense that
 \begin{align}
 \label{eq:PsiNormalization}
 \left\Vert\Psi(\alpha,\bx,\bx_{\rm c})\right\Vert_{2}=1.
 \end{align}
}
\item{
The sequence of such functions
 \begin{align}
 \label{eq:SingularSequence}
 \Psi(\alpha_{n},\bx,\bx_{\rm c}),\;\;\;\;\; \alpha_{n}>\alpha_{n-1},\;\;\;\;\;n=1,2,\dots
 \end{align}
does not have a convergent subsequence.
}
\item{Its Fourier transform is given by
 \begin{align}
 \label{eq:PsiFourier}
 \tilde\Psi(\alpha,\bk,\bx_{\rm c})=-i\left(\frac{2}{3}\right)^{1/2}\pi^{-3/4}\alpha^{-5/4}\bk 
\exp\left(-\frac{1}{2\alpha}\vert\bk\vert^{2}-i\bk\cdot\bx_{\rm c}\right).
 \end{align}
}
\item{It is a generator of the square-root of the Dirac delta-function, i.e., with any bounded 
continuous function $f(\bx)$ its square has the sifting property
 \begin{align}
 \label{eq:PsiSquare}
 \lim\limits_{\alpha\rightarrow\infty}\int\limits_{\bx\in{\mathbb R}^{3}} 
 f(\bx)\left\vert\Psi(\alpha,\bx,\bx_{\rm c})\right\vert^{2}\,{\rm d}\bx = f(\bx_{\rm c}).
 \end{align}
}
\item{It is orthogonal to bounded vector-valued functions, i.e., for any $\left\vert\bV(\bx)\right\vert<\infty$, $\bx\in{\mathbb R}^{3}$,
 \begin{align}
 \label{eq:PsiOrthogonalBounded}
 \lim\limits_{\alpha\rightarrow\infty}\left\langle\bV,\Psi\right\rangle = 
 \lim\limits_{\alpha\rightarrow\infty}\int\limits_{\bx\in{\mathbb R}^{3}}
 \Psi^{\rm T}(\alpha,\bx,\bx_{\rm c})\bV(\bx)\,{\rm d}\bx = 0.
 \end{align}
}
\item{It is `invisible' to weakly singular operators with finite spatial support, i.e.
 \begin{align}
 \label{eq:PsiWeaklySingular}
 \lim\limits_{\alpha\rightarrow\infty}
 \left\Vert\int\limits_{\bx\in D}
 \frac{{\mathbb K}(\bx,\bx')}{\vert\bx-\bx'\vert^{\beta}}\Psi(\alpha,\bx,\bx_{\rm c})\,{\rm d}\bx\right\Vert_{2} = 0,
 \end{align}
where ${\mathbb K}(\bx,\bx')$, $\bx,\bx'\in{\mathbb R}^{3}$ is a bounded tensor-valued function, $\beta<3$, and 
the norm is defined over the spatial support $D$. 
}
\end{enumerate}
\end{theorem} 
\begin{proof}

1. To prove the normalization property we simply compute
 \begin{align}
 \label{eq:PsiTwoNormProof}
 \begin{split}
 \left\Vert\Psi(\alpha,\bx,\bx_{\rm c})\right\Vert_{2}^{2}&=
 \frac{2}{3}\pi^{-3/2}\alpha^{5/2}\int\limits_{\bx\in{\mathbb R}^{3}}\vert\bx-\bx_{\rm c}\vert^{2}e^{-\alpha\vert\bx-\bx_{\rm c}\vert^{2}}\,{\rm d}\bx
 \\
 &= \frac{2}{3}\pi^{-3/2}\alpha^{5/2}\int\limits_{r=0}^{\infty}\int\limits_{\theta=0}^{\pi}\int\limits_{\varphi=0}^{2\pi}r^{4}e^{-\alpha r^{2}}
 \,\sin\theta{\rm d}\varphi {\rm d}\theta{\rm d}r
 \\
 &=\frac{2}{3}\pi^{-3/2}\alpha^{5/2} 4\pi \frac{(4-1)!!}{2(2\alpha)^{2}}\left(\frac{\pi}{\alpha}\right)^{1/2}=1,
 \end{split}
 \end{align}
where we have used the following standard integral:
 \begin{align}
 \label{eq:StandardEven}
 \int\limits_{0}^{\infty}r^{2n}e^{-pr^{2}}\,{\rm d}r=
 \frac{(2n-1)!!}{2(2p)^{n}}\sqrt{\frac{\pi}{p}},\;\;\;\;p>0,\;\;\;\;n=0,1,2,\dots
 \end{align}

2. Suppose that there is a subsequence $\Psi(\alpha_{n},\bx,\bx_{\rm c})$, where $\alpha_{n}>\alpha_{n-1}$, $n=1,2,\dots$, 
which converges in norm. Then, for any $\epsilon>0$ there exists $N$ such that for all $m,n>N$ we have
 \begin{align}
 \label{eq:Convergent}
 \begin{split}
 \left\Vert \Psi(\alpha_{m},\bx,\bx_{\rm c}) - \Psi(\alpha_{n},\bx,\bx_{\rm c})\right\Vert^{2}\le \epsilon.
 \end{split}
 \end{align}
However, in our case we obtain
 \begin{align}
 \label{eq:NonConvergent}
 \begin{split}
 &\left\Vert \Psi(\alpha_{m},\bx,\bx_{\rm c}) - \Psi(\alpha_{n},\bx,\bx_{\rm c})\right\Vert^{2} 
 = 
 \int\limits_{\bx\in{\mathbb R}^{3}}\left\vert\Psi(\alpha_{m},\bx,\bx_{\rm c}) - \Psi(\alpha_{n},\bx,\bx_{\rm c})\right\vert^{2}\,{\rm d}\bx
 \\
 &=\int\limits_{\bx\in{\mathbb R}^{3}}\left\vert\Psi(\alpha_{m},\bx,\bx_{\rm c})\right\vert^{2}\,{\rm d}\bx
 +\int\limits_{\bx\in{\mathbb R}^{3}}\left\vert\Psi(\alpha_{n},\bx,\bx_{\rm c})\right\vert^{2}\,{\rm d}\bx
 \\
 &-2\int\limits_{\bx\in{\mathbb R}^{3}}\Psi^{T}(\alpha_{m},\bx,\bx_{\rm c})\Psi(\alpha_{n},\bx,\bx_{\rm c})\,{\rm d}\bx
 \\
 &=2 - 2 \frac{\alpha_{m}^{5/4}\alpha_{n}^{5/4}}{(\alpha_{m}+\alpha_{n})^{5/2}}
 \int\limits_{\bx\in{\mathbb R}^{3}}\left\vert\Psi(\alpha_{m}+\alpha_{n},\bx,\bx_{\rm c})\right\vert^{2}\,{\rm d}\bx
 \\
 &=2-2\frac{\alpha_{m}^{5/4}\alpha_{n}^{5/4}}{(\alpha_{m}+\alpha_{n})^{5/2}}.
 \end{split}
 \end{align}
If we now fix $n>N$ and let $m>N$ go to infinity, then the last term tends to zero leaving us with a constant (two), which, 
obviously, cannot be made smaller than an arbitrary $\epsilon>0$. This proves that there are no convergent 
subsequences.

3. The Fourier transform is obtained by a direct computation as follows:
 \begin{align}
 \label{eq:PsiFourierProof}
 \begin{split}
 &\tilde\Psi(\alpha,\bk,\bx_{\rm c})=(2\pi)^{-3/2}\int\limits_{\bx\in{\mathbb R}^{3}}
 \Psi(\alpha,\bx,\bx_{\rm c})\exp({-i\bk\cdot\bx})\,{\rm d}\bx
 \\
 &=(2\pi)^{-3/2}\left(\frac{2}{3}\right)^{1/2}\pi^{-3/4}\alpha^{5/4}\int\limits_{\bx\in{\mathbb R}^{3}}
 (\bx-\bx_{\rm c})\exp\left({-\frac{\alpha}{2}\vert\bx-\bx_{\rm c}\vert^{2}-i\bk\cdot\bx}\right)\,{\rm d}\bx
 \\
 &=(2\pi)^{-3/2}\left(\frac{2}{3}\right)^{1/2}\pi^{-3/4}\alpha^{5/4}\exp({-i\bk\cdot\bx_{\rm c}})
 \int\limits_{\by\in{\mathbb R}^{3}}
 \by \exp\left({-\frac{\alpha}{2}\vert\by\vert^{2}-i\bk\cdot\by}\right)\,{\rm d}\by
 \\
 &=(2\pi)^{-3/2}\left(\frac{2}{3}\right)^{1/2}\pi^{-3/4}\alpha^{5/4}\exp({-i\bk\cdot\bx_{\rm c}}) (i\nabla_{\bk})
 \int\limits_{\by\in{\mathbb R}^{3}}\exp\left({-\frac{\alpha}{2}\vert\by\vert^{2}-i\bk\cdot\by}\right)\,{\rm d}\by
 \\
 &=(2\pi)^{-3/2}\left(\frac{2}{3}\right)^{1/2}\pi^{-3/4}\alpha^{5/4}\exp({-i\bk\cdot\bx_{\rm c}}) (i\nabla_{\bk})
 \int\limits_{-\infty}^{\infty}\exp\left(-\frac{\alpha}{2}y_{1}^{2}-ik_{1}y_{1}\right)\,{\rm d}y_{1}
 \\
 &\times\int\limits_{-\infty}^{\infty}\exp\left(-\frac{\alpha}{2}y_{2}^{2}-ik_{2}y_{2}\right)\,{\rm d}y_{2}
 \times\int\limits_{-\infty}^{\infty}\exp\left(-\frac{\alpha}{2}y_{3}^{2}-ik_{3}y_{3}\right)\,{\rm d}y_{3}.
 \end{split}
 \end{align}
Each of the one-dimensional integrals above gives 
 \begin{align}
 \label{eq:1DFourier}
 \begin{split}
 &\int\limits_{-\infty}^{\infty}\exp\left(-\frac{\alpha}{2}y_{n}^{2}-ik_{n}y_{n}\right)\,{\rm d}y_{n} = 
 \int\limits_{-\infty}^{\infty}\exp\left[-\frac{\alpha}{2}\left(y_{n}^{2}+i\frac{2}{\alpha}k_{n}y_{n}\right)\right]\,{\rm d}y_{n}
 \\
 &=\int\limits_{-\infty}^{\infty}
 \exp\left[-\frac{\alpha}{2}\left(y_{n}^{2}+2y_{n}\left(i\frac{k_{n}}{\alpha}\right)+\left(i\frac{k_{n}}{\alpha}\right)^{2}
 -\left(i\frac{k_{n}}{\alpha}\right)^{2}\right)\right]\,{\rm d}y_{n}
 \\
 &=\exp\left(-\frac{1}{2\alpha}k_{n}^{2}\right)\int\limits_{-\infty}^{\infty}
 \exp\left[-\frac{\alpha}{2}\left(y_{n}+\frac{i}{\alpha}k_{n}\right)^{2}\right]\,{\rm d}y_{n}
 \\
 &=\exp\left(-\frac{1}{2\alpha}k_{n}^{2}\right)\int\limits_{-\infty}^{\infty}
 \exp\left[-\pi\left(\sqrt{\frac{\alpha}{2\pi}}y_{n}+\frac{i}{\sqrt{2\pi\alpha}}k_{n}\right)^{2}\right]\,{\rm d}y_{n}
 \\
 &=\left(\frac{2\pi}{\alpha}\right)^{1/2}\exp\left(-\frac{1}{2\alpha}k_{n}^{2}\right)\int\limits_{-\infty+ib}^{\infty+ib}
 \exp\left[-\pi (z+ib)^{2}\right]\,{\rm d}(z+ib)
 \\
 &=\left(\frac{2\pi}{\alpha}\right)^{1/2}\exp\left(-\frac{1}{2\alpha}k_{n}^{2}\right).
 \end{split}
 \end{align}
Substituting (\ref{eq:1DFourier}) in (\ref{eq:PsiFourierProof}) we arrive at our result
 \begin{align}
 \label{eq:PsiFourierNabla}
 \begin{split}
 &\tilde\Psi(\alpha,\bk,\bx_{\rm c})=
 \\
 &i (2\pi)^{-3/2}\left(\frac{2}{3}\right)^{1/2}\pi^{-3/4}\alpha^{5/4}\exp({-i\bk\cdot\bx_{\rm c}}) 
 \nabla_{\bk}\left[\left(\frac{2\pi}{\alpha}\right)^{3/2}\exp\left(-\frac{1}{2\alpha}\vert\bk\vert^{2}\right)\right]
 \\
 &=- i (2\pi)^{-3/2}\left(\frac{2}{3}\right)^{1/2}\pi^{-3/4}\alpha^{5/4}\left(\frac{2\pi}{\alpha}\right)^{3/2}\frac{1}{2\alpha}
 \exp\left({-\frac{1}{2\alpha}\vert\bk\vert^{2}-i\bk\cdot\bx_{\rm c}}\right) \nabla_{\bk}\vert\bk\vert^{2}
 \\
 &=-i\left(\frac{2}{3}\right)^{1/2}\pi^{-3/4}\alpha^{-5/4}\bk 
\exp\left(-\frac{1}{2\alpha}\vert\bk\vert^{2}-i\bk\cdot\bx_{\rm c}\right).
 \end{split}
 \end{align}
Note that with this choice of the Fourier transform we also have
 \begin{align}
 \label{eq:PsiFourierNormalization}
 \left\Vert\tilde{\Psi}(\alpha,\bk,\bx_{\rm c})\right\Vert_{2}=1.
 \end{align}

4. To prove the sifting property we split the integration domain in two parts, i.e.,
 \begin{align}
 \label{eq:PsiSifting}
 \begin{split}
 \lim\limits_{\alpha\rightarrow\infty} 
 &\int\limits_{\bx\in{\mathbb R}^{3}}f(\bx)\left\vert\Psi(\alpha,\bx,\bx_{\rm c})\right\vert^{2}\,{\rm d}\bx 
 \\
 &=\lim\limits_{\alpha\rightarrow\infty}
 \left[ 
 \int\limits_{\bx\in{\mathbb R}^{3}\setminus V(\delta)}f(\bx)\left\vert\Psi(\alpha,\bx,\bx_{\rm c})\right\vert^{2}\,{\rm d}\bx
 +\int\limits_{\bx\in V(\delta)}f(\bx)\left\vert\Psi(\alpha,\bx,\bx_{\rm c})\right\vert^{2}\,{\rm d}\bx
 \right]
 \\
 &=\lim\limits_{\alpha\rightarrow\infty}
 \int\limits_{\bx\in{\mathbb R}^{3}\setminus V(\delta)}f(\bx)\left\vert\Psi(\alpha,\bx,\bx_{\rm c})\right\vert^{2}\,{\rm d}\bx
 +\lim\limits_{\alpha\rightarrow\infty}
 \int\limits_{\bx\in V(\delta)}f(\bx)\left\vert\Psi(\alpha,\bx,\bx_{\rm c})\right\vert^{2}\,{\rm d}\bx,
 \end{split}
 \end{align}
where $V(\delta)$ is some volume surrounding the point $\bx_{\rm c}$.
Now we shall choose $V(\delta)$ in such a way that  
the first (outer) term in the expression above gives zero. We start by considering a general case where 
$\bx\in V(\delta)$, if $\vert\bx-\bx_{\rm c}\vert\le\delta$, and $\delta$ is some function of $\alpha$. Then,
 \begin{align}
 \label{eq:SiftingOuterPart}
 \begin{split}
 \lim\limits_{\alpha\rightarrow\infty}
 &\int\limits_{\bx\in{\mathbb R}^{3}\setminus V(\delta)}f(\bx)\left\vert\Psi(\alpha,\bx,\bx_{\rm c})\right\vert^{2}\,{\rm d}\bx
 \\
 &\le\max\limits_{\bx\in{\mathbb R}^{3}}\vert f(\bx) \vert
 \lim\limits_{\alpha\rightarrow\infty} \int\limits_{\bx\in{\mathbb R}^{3}\setminus V(\delta)}\left\vert\Psi(\alpha,\bx,\bx_{\rm c})\right\vert^{2}\,{\rm d}\bx
 \\
 &=\max\limits_{\bx\in{\mathbb R}^{3}}\vert f(\bx) \vert \lim\limits_{\alpha\rightarrow\infty}
 \frac{2}{3}\pi^{-3/2}\alpha^{5/2}\int\limits_{r=\delta}^{\infty}\int\limits_{\theta=0}^{\pi}\int\limits_{\varphi=0}^{2\pi}r^{4}e^{-\alpha r^{2}}
 \,\sin\theta{\rm d}\varphi{\rm d}\theta {\rm d}r
 \\
 &=\frac{8\pi}{3}\pi^{-3/2}\max\limits_{\bx\in{\mathbb R}^{3}}\vert f(\bx) \vert \lim\limits_{\alpha\rightarrow\infty}
 \alpha^{5/2}\int\limits_{r=\delta}^{\infty}r^{4}e^{-\alpha r^{2}}\,{\rm d}r.
 \end{split}
 \end{align}
Successive integration by parts gives
 \begin{align}
 \label{eq:ByParts}
 \begin{split}
 \int\limits_{r=\delta}^{\infty}r^{4}e^{-\alpha r^{2}}\,{\rm d}r = 
 \frac{3}{4\alpha^{2}}\int\limits_{r=\delta}^{\infty}e^{-\alpha r^{2}}\,{\rm d}r +
 \left(\frac{\delta^{3}}{2\alpha}+\frac{3\delta}{4\alpha^{2}}\right)e^{-\alpha\delta^{2}}.
 \end{split}
 \end{align}
 Using this result we continue to analyze the upper bound of (\ref{eq:SiftingOuterPart}) as follows:
 \begin{align}
 \label{eq:DeltaToAlpha}
 \begin{split}
 \lim\limits_{\alpha\rightarrow\infty} &\int\limits_{\bx\in{\mathbb R}^{3}\setminus V(\delta)}\left\vert\Psi(\alpha,\bx,\bx_{\rm c})\right\vert^{2}\,{\rm d}\bx=
 \lim\limits_{\alpha\rightarrow\infty}
 \frac{8\pi}{3}\pi^{-3/2}\alpha^{5/2}\int\limits_{r=\delta}^{\infty}r^{4}e^{-\alpha r^{2}}\,{\rm d}r 
 \\
 &=\frac{8\pi}{3}\pi^{-3/2}\lim\limits_{\alpha\rightarrow\infty}\alpha^{5/2}\left[\frac{3}{4\alpha^{2}}\int\limits_{r=\delta}^{\infty}e^{-\alpha r^{2}}\,{\rm d}r +
 \left(\frac{\delta^{3}}{2\alpha}+\frac{3\delta}{4\alpha^{2}}\right)e^{-\alpha\delta^{2}}\right]
 \\
 &=\frac{2}{\sqrt{\pi}}\lim\limits_{\alpha\rightarrow\infty}\alpha^{1/2}\int\limits_{r=\delta}^{\infty}e^{-\alpha r^{2}}\,{\rm d}r +
 \lim\limits_{\alpha\rightarrow\infty}
 \frac{8}{3\sqrt{\pi}}\left(\frac{1}{2}\delta^{3}\alpha^{2/3}+\frac{3}{4}\delta\alpha^{1/2}\right)e^{-\alpha\delta^{2}}.
 \end{split}
 \end{align}
Now, choosing, for example, $\delta=\alpha^{-1/3}$, we arrive at
 \begin{align}
 \label{eq:MinOneThirdAlpha}
 \begin{split}
 \lim\limits_{\alpha\rightarrow\infty} &\int\limits_{\bx\in{\mathbb R}^{3}\setminus V(\delta)}\left\vert\Psi(\alpha,\bx,\bx_{\rm c})\right\vert^{2}\,{\rm d}\bx
 \\
 &=\frac{2}{\sqrt{\pi}}\lim\limits_{\alpha\rightarrow\infty}\int\limits_{r\sqrt{\alpha}=\alpha^{1/6}}^{\infty}e^{- (r\sqrt{\alpha})^{2}}\,{\rm d} (r\sqrt{\alpha})
 +\lim\limits_{\alpha\rightarrow\infty}
 \frac{8}{3\sqrt{\pi}}\left(\frac{1}{2}\alpha^{1/2}+\frac{3}{4}\alpha^{1/6}\right)e^{-\alpha^{1/3}}
 \\
 &\le\frac{2}{\sqrt{\pi}}\lim\limits_{\alpha\rightarrow\infty}\int\limits_{z=\alpha^{1/6}}^{\infty}e^{- z}\,{\rm d} z 
 = \frac{2}{\sqrt{\pi}}\lim\limits_{\alpha\rightarrow\infty}e^{-\alpha^{1/6}}= 0.
 \end{split}
 \end{align}
Hence, with this particular choice of $V(\delta)$ the first (outer) term in (\ref{eq:PsiSifting}) is zero. Now we shall 
use the same $V(\delta)$ in the second (inner) term. Taking into account that $f(\bx)$ is a continuous function, and that
with our choice of $\delta$ the integration volume $V(\delta)$ tends to the point $\bx_{c}$, we can apply 
the mean-value theorem, i.e.,
 \begin{align}
 \label{eq:PsiInner}
 \begin{split}
 \lim\limits_{\alpha\rightarrow\infty}
 \int\limits_{\bx\in V(\delta)}f(\bx)\left\vert\Psi(\alpha,\bx,\bx_{\rm c})\right\vert^{2}\,{\rm d}\bx
 &=\lim\limits_{\alpha\rightarrow\infty}f(\bx_{\alpha})
 \int\limits_{\bx\in V(\delta)}\left\vert\Psi(\alpha,\bx,\bx_{\rm c})\right\vert^{2}\,{\rm d}\bx
 \\
 &=f(\bx_{\rm c})\lim\limits_{\alpha\rightarrow\infty}
 \int\limits_{\bx\in V(\delta)}\left\vert\Psi(\alpha,\bx,\bx_{\rm c})\right\vert^{2}\,{\rm d}\bx,
 \end{split}
 \end{align}
where $\bx_{\alpha}\in V(\delta)$, and $\bx_{\alpha}\rightarrow\bx_{\rm c}$ as $\alpha\rightarrow\infty$.
Thus, to prove the sifting property it remains to show that
 \begin{align}
 \label{eq:InnerOne}
 \begin{split}
 \lim\limits_{\alpha\rightarrow\infty}
 &\int\limits_{\bx\in V(\delta)}\left\vert\Psi(\alpha,\bx,\bx_{\rm c})\right\vert^{2}\,{\rm d}\bx
 =\frac{8}{3\sqrt{\pi}} \lim\limits_{\alpha\rightarrow\infty} \alpha^{5/2}
 \int\limits_{r=0}^{\alpha^{-1/3}}r^{4}e^{-\alpha r^{2}}\,{\rm d}r
 \\
 &=\frac{8}{3\sqrt{\pi}} \lim\limits_{\alpha\rightarrow\infty} 
 \int\limits_{r\sqrt{\alpha}=0}^{\alpha^{1/6}}(r\sqrt{\alpha})^{4}e^{-(r\sqrt{\alpha})^{2}}\,{\rm d}(r\sqrt{\alpha})
 =\frac{8}{3\sqrt{\pi}} \lim\limits_{\alpha\rightarrow\infty}
 \int\limits_{z=0}^{\alpha^{1/6}}z^{4}e^{-z^{2}}\,{\rm d}z
 \\
 &=\frac{8}{3\sqrt{\pi}} \int\limits_{z=0}^{\infty}z^{4}e^{-z^{2}}\,{\rm d}z = 1.
 \end{split}
 \end{align}
Finally, we remark that the above proof holds with any $\delta=\alpha^{-1/m}$, where $m$ is an integer $m\ge 3$.

5. We prove the orthogonality property by considering the
absolute value of the dot-product, i.e.,
 \begin{align}
 \label{eq:OrthogonalityProof}
 \begin{split}
 &\left\vert\lim\limits_{\alpha\rightarrow\infty}
 \left\langle\bV(\bx),\Psi(\alpha,\bx,\bx_{\rm c})\right\rangle\right\vert
 \\
 &=\left\vert\lim\limits_{\alpha\rightarrow\infty}
 \sqrt{\frac{2}{3}}\pi^{-3/4}\alpha^{5/4}
 \int\limits_{r=0}^{\infty}\int\limits_{\theta=0}^{\pi}\int\limits_{\varphi=0}^{2\pi}
 \bTheta^{\rm T}(\theta,\varphi)\bV(r,\theta,\varphi)r^{3}e^{-\frac{\alpha}{2} r^{2}}
 \,\sin\theta{\rm d}\varphi {\rm d}\theta{\rm d}r\right\vert
 \\
 &\le C \max\limits_{\bx\in{\mathbb R}^{3}}\left\vert\bTheta^{\rm T}(\theta,\varphi)\bV(\bx)\right\vert
 \lim\limits_{\alpha\rightarrow\infty}\alpha^{5/4}\int\limits_{r=0}^{\infty}r^{3}e^{-\frac{\alpha}{2} r^{2}}\,{\rm d}r
 \\
 &=C \max\limits_{\bx\in{\mathbb R}^{3}}\left\vert\bTheta^{\rm T}(\theta,\varphi)\bV(\bx)\right\vert
 \lim\limits_{\alpha\rightarrow\infty}\alpha^{5/4} \frac{1}{2(\alpha/2)^{2}}
 \\
 &=2 C \max\limits_{\bx\in{\mathbb R}^{3}}\left\vert\bTheta^{\rm T}(\theta,\varphi)\bV(\bx)\right\vert
 \lim\limits_{\alpha\rightarrow\infty}\alpha^{-3/4}=0,
 \end{split}
 \end{align}
where $0<C<\infty$, $\bTheta=(\bx-\bx_{\rm c})/\vert\bx-\bx_{\rm c}\vert$, and the following standard
integral was used:
 \begin{align}
 \label{eq:StandardOdd}
 \int\limits_{0}^{\infty}r^{2n+1}e^{-pr^{2}}\,{\rm d}r=
 \frac{n!}{2p^{n+1}},\;\;\;\;p>0,\;\;\;\;n=0,1,2,\dots
 \end{align}

6. In the case of a weakly singular integral operator with finite spatial support $D$ we proceed as follows:
 \begin{align}
 \label{eq:PsiWeaklySingularProof}
 \begin{split}
 &\lim\limits_{\alpha\rightarrow\infty}
 \left\Vert\;\int\limits_{\bx\in D}
 \frac{{\mathbb K}(\bx,\bx')}{\vert\bx-\bx'\vert^{\beta}}\Psi(\alpha,\bx,\bx_{\rm c})\,{\rm d}\bx\right\Vert_{2}^{2} 
 \\
 &=\lim\limits_{\alpha\rightarrow\infty}
 \int\limits_{\bx'\in D}\left\vert 
 \;
 \int\limits_{\bx\in D}
 \frac{{\mathbb K}(\bx,\bx')}{\vert\bx-\bx'\vert^{\beta}}\Psi(\alpha,\bx,\bx_{\rm c})\,{\rm d}\bx
 \right\vert^{2}\,{\rm d}\bx'
 \\
 &=\lim\limits_{\alpha\rightarrow\infty}
 \int\limits_{\bx'\in D}\left\vert 
 \;
 \int\limits_{\bx\in D}
 \frac{{\mathbb K}(\bx,\bx')}{\vert\bx-\bx'\vert^{\beta}}\left[\Psi_{2}(\alpha,\bx,\bx_{\rm c})
 +\Psi_{1}(\alpha,\bx,\bx_{\rm c})\right]\,{\rm d}\bx
 \right\vert^{2}\,{\rm d}\bx'
 \\
 &\le\lim\limits_{\alpha\rightarrow\infty}
 \int\limits_{\bx'\in D}\left\vert 
 \;
 \int\limits_{\bx\in D}
 \frac{{\mathbb K}(\bx,\bx')}{\vert\bx-\bx'\vert^{\beta}}\Psi_{1}(\alpha,\bx,\bx_{\rm c})\,{\rm d}\bx
 \right\vert^{2}\,{\rm d}\bx'
 \\
 &+\lim\limits_{\alpha\rightarrow\infty}
 \int\limits_{\bx'\in D}\left\vert 
 \;
 \int\limits_{\bx\in D}
 \frac{{\mathbb K}(\bx,\bx')}{\vert\bx-\bx'\vert^{\beta}}\Psi_{2}(\alpha,\bx,\bx_{\rm c})\,{\rm d}\bx
 \right\vert^{2}\,{\rm d}\bx',
 \end{split}
 \end{align}
where the original function $\Psi$ is split into two complementary parts with respect to a small volume around $\bx_{\rm c}$
in such a way that:
 \begin{align}
 \label{eq:SplitPsi}
 \begin{split}
 &\Psi(\alpha,\bx,\bx_{\rm c})=\Psi_{1}(\alpha,\bx,\bx_{\rm c})+\Psi_{2}(\alpha,\bx,\bx_{\rm c}),
 \\
 &\Psi_{1}(\alpha,\bx,\bx_{\rm c})=0, \;\;\;\;\;\;\bx\in {\mathbb R}^{3}\setminus V(\delta),
 \\
 &\Psi_{2}(\alpha,\bx,\bx_{\rm c})=0, \;\;\;\;\;\;\bx\in V(\delta).
 \end{split}
 \end{align}
The last integral in (\ref{eq:PsiWeaklySingularProof}) is estimated like this
 \begin{align}
 \label{eq:Psi2Part}
 \begin{split}
 &\lim\limits_{\alpha\rightarrow\infty}
 \int\limits_{\bx'\in D}\left\vert 
 \;
 \int\limits_{\bx\in D}
 \frac{{\mathbb K}(\bx,\bx')}{\vert\bx-\bx'\vert^{\beta}}\Psi_{2}(\alpha,\bx,\bx_{\rm c})\,{\rm d}\bx
 \right\vert^{2}\,{\rm d}\bx'
 \\
 & = \lim\limits_{\alpha\rightarrow\infty} \left\Vert K \Psi_{2}\right\Vert^{2} 
 \le
 \left\Vert K \right\Vert^{2} \lim\limits_{\alpha\rightarrow\infty} \left\Vert \Psi_{2}\right\Vert^{2} =0,
 \end{split}
 \end{align}
where we have used the fact that the norm of a weakly singular operator on $D$ is bounded, and the previously 
derived property (\ref{eq:DeltaToAlpha}) -- (\ref{eq:MinOneThirdAlpha}). This means that we take $\delta=\alpha^{-1/m}$,
$m\ge 3$. The remaining integral in (\ref{eq:PsiWeaklySingularProof}) 
requires considerably more work. We shall split the domain of integration over $\bx'$ in two parts using yet another small volume $V(\delta')$ 
surrounding the point $\bx_{\rm c}$. Then, taking (\ref{eq:SplitPsi}) into account, we obtain
 \begin{align}
 \label{eq:Psi1Part}
 \begin{split}
 &\lim\limits_{\alpha\rightarrow\infty}
 \int\limits_{\bx'\in D}\left\vert 
 \;
 \int\limits_{\bx\in D}
 \frac{{\mathbb K}(\bx,\bx')}{\vert\bx-\bx'\vert^{\beta}}\Psi_{1}(\alpha,\bx,\bx_{\rm c})\,{\rm d}\bx
 \right\vert^{2}\,{\rm d}\bx' 
 \\&=\lim\limits_{\alpha\rightarrow\infty}
 \int\limits_{\bx'\in D\setminus V(\delta')}\left\vert 
 \;\int\limits_{\bx\in V(\delta)}
 \frac{{\mathbb K}(\bx,\bx')}{\vert\bx-\bx'\vert^{\beta}}\Psi_{1}(\alpha,\bx,\bx_{\rm c})\,{\rm d}\bx
 \right\vert^{2}\,{\rm d}\bx'
 \\&+\lim\limits_{\alpha\rightarrow\infty}
 \int\limits_{\bx'\in V(\delta')}\left\vert 
 \;\int\limits_{\bx\in V(\delta)}
 \frac{{\mathbb K}(\bx,\bx')}{\vert\bx-\bx'\vert^{\beta}}\Psi_{1}(\alpha,\bx,\bx_{\rm c})\,{\rm d}\bx
 \right\vert^{2}\,{\rm d}\bx'
 \end{split}
 \end{align}
Proceeding with the first of the above integrals we apply the Caushy-Schwartz inequality and arrive at
 \begin{align}
 \label{eq:Psi1FirstPart}
 \begin{split}
 &\lim\limits_{\alpha\rightarrow\infty}
 \int\limits_{\bx'\in D\setminus V(\delta')}\left\vert 
 \;\int\limits_{\bx\in V(\delta)}
 \frac{{\mathbb K}(\bx,\bx')}{\vert\bx-\bx'\vert^{\beta}}\Psi_{1}(\alpha,\bx,\bx_{\rm c})\,{\rm d}\bx
 \right\vert^{2}\,{\rm d}\bx'
 \\
 &\le\lim\limits_{\alpha\rightarrow\infty}
 \int\limits_{\bx'\in D\setminus V(\delta')}
 \int\limits_{\bx\in V(\delta)}
 \frac{\left\vert{\mathbb K}(\bx,\bx')\bTheta\right\vert^{2}}{\vert\bx-\bx'\vert^{2\beta}}\,{\rm d}\bx
 \int\limits_{\bx\in V(\delta)}
 \left\vert\Psi_{1}(\alpha,\bx,\bx_{\rm c})\right\vert^{2}\,{\rm d}\bx
 \,{\rm d}\bx'
 \\
 &\le \max\limits_{\bx,\bx'\in D}\left\vert{\mathbb K}(\bx,\bx')\bTheta\right\vert^{2}
 \\
 &\times
 \lim\limits_{\alpha\rightarrow\infty}
 \int\limits_{\bx'\in D\setminus V(\delta')}
 \int\limits_{\bx\in V(\delta)}
 \frac{1}{\vert\bx-\bx'\vert^{2\beta}}\,{\rm d}\bx\,{\rm d}\bx'
 \int\limits_{\bx\in V(\delta)}
 \left\vert\Psi_{1}(\alpha,\bx,\bx_{\rm c})\right\vert^{2}\,{\rm d}\bx
 \end{split}
 \end{align}
Now, if we choose the spherical volume $V(\delta')$ with radius $\delta'$  to be larger than the spherical 
volume $V(\delta)$ with radius $\delta$, then for $\bx\in V(\delta)$ and $\bx'\in D\setminus V(\delta')$, 
the distance factor $\vert\bx-\bx'\vert$ will be bounded from below by the difference 
of the radii of the two volumes. In the estimate  (\ref{eq:Psi2Part}) we have used 
$\delta=\alpha^{-1/m}$, $m\ge 3$. Hence, for sufficiently large $\alpha$ we can 
choose $\delta'=\delta^{1/n}=\alpha^{-1/(nm)}$ with integer $n\ge 2$.
In this case we have
 \begin{align}
 \label{eq:TwoRadii}
 \begin{split}
 &\lim\limits_{\alpha\rightarrow\infty}
 \int\limits_{\bx'\in D\setminus V(\delta')}
 \int\limits_{\bx\in V(\delta)}
 \frac{1}{\vert\bx-\bx'\vert^{2\beta}}\,{\rm d}\bx
 \int\limits_{\bx\in V(\delta)}
 \left\vert\Psi_{1}(\alpha,\bx,\bx_{\rm c})\right\vert^{2}\,{\rm d}\bx
 \,{\rm d}\bx'
 \\
 &\le \lim\limits_{\alpha\rightarrow\infty}
 \max\limits_{\bx\in V(\delta), \bx'\in D\setminus V(\delta')}\frac{1}{\vert\bx-\bx'\vert^{2\beta}}
 \int\limits_{\bx'\in D\setminus V(\delta')}
 \int\limits_{\bx\in V(\delta)}\;{\rm d}\bx\,{\rm d}\bx'
 \int\limits_{\bx\in V(\delta)}
 \left\vert\Psi_{1}(\alpha,\bx,\bx_{\rm c})\right\vert^{2}\,{\rm d}\bx
 \\
 &\le C \lim\limits_{\alpha\rightarrow\infty}
 \frac{\delta^{3}}{\left(\delta'-\delta\right)^{2\beta}}
 = C \lim\limits_{\alpha\rightarrow\infty}
 \frac{\delta^{3}}{\left(\delta^{1/n}-\delta\right)^{2\beta}}=C \lim\limits_{\alpha\rightarrow\infty}
 \frac{\delta^{3-1/n}}{\left(1-\delta^{1-1/n}\right)^{2\beta}}
 \\
 &=C \lim\limits_{\alpha\rightarrow\infty}
 \frac{\alpha^{(1-3n)/(nm)}}{\left(1-\alpha^{(1-n)/(nm)}\right)^{2\beta}}
 =0,
 \end{split}
 \end{align}
i.e., the first of the two integrals in (\ref{eq:Psi1Part}) is zero. Applying the Caushy-Schwartz inequality we 
estimate the last integral in
(\ref{eq:Psi1Part}) as follows:
 \begin{align}
 \label{eq:Psi1SecondPart}
 \begin{split}
 &\lim\limits_{\alpha\rightarrow\infty}
 \int\limits_{\bx'\in V(\delta')}\left\vert 
 \;\int\limits_{\bx\in V(\delta)}
 \frac{{\mathbb K}(\bx,\bx')}{\vert\bx-\bx'\vert^{\beta}}\Psi_{1}(\alpha,\bx,\bx_{\rm c})\,{\rm d}\bx
 \right\vert^{2}\,{\rm d}\bx'
 \\
 &\le K \lim\limits_{\alpha\rightarrow\infty}
 \int\limits_{\bx'\in V(\delta')}
 \int\limits_{\bx\in V(\delta)} \frac{1}{\vert\bx-\bx'\vert^{\beta}}\,{\rm d}\bx
 \int\limits_{\bx\in V(\delta)} \frac{\vert\Psi_{1}(\alpha,\bx,\bx_{\rm c})\vert^{2}}{\vert\bx-\bx'\vert^{\beta}}\,{\rm d}\bx
 \,{\rm d}\bx'
 \\
 & \le L \lim\limits_{\alpha\rightarrow\infty} (\delta')^{3-\beta}
 \int\limits_{\bx'\in V(\delta')}
 \int\limits_{\bx\in V(\delta)} \frac{\vert\Psi_{1}(\alpha,\bx,\bx_{\rm c})\vert^{2}}{\vert\bx-\bx'\vert^{\beta}}\,{\rm d}\bx
 \,{\rm d}\bx'
 \\
 & \le M \lim\limits_{\alpha\rightarrow\infty} (\delta')^{2(3-\beta)}
 \int\limits_{\bx\in V(\delta)} \vert\Psi_{1}(\alpha,\bx,\bx_{\rm c})\vert^{2}\,{\rm d}\bx
 \\
 & = N \lim\limits_{\alpha\rightarrow\infty} (\delta')^{6-2\beta} = N \lim\limits_{\alpha\rightarrow\infty} \alpha^{-(6-2\beta)/(nm)}=0,
 \end{split}
 \end{align}
where $\delta'=\alpha^{-1/(nm)}$, $m\ge 3,\,n\ge 2$, while $0<\beta<3$ by the conditions of the theorem. 
Thus, we have shown that (\ref{eq:PsiWeaklySingularProof}) is, indeed, zero.
\end{proof}

\section{Electromagnetic singular modes}
%%%%%%%%%%%%%%%%%%%%%%%%%%%%%%%%%%%%%%%%%%%%%%%%%%%%%%%%%%%%%%%%%%%%%%%%%%%%%%
%%%%%%%%%%%%%%%%%%%%%%%%%%%%%%%%%%%%%%%%%%%%%%%%%%%%%%%%%%%%%%%%%%%%%%%%%%%%%%
Consider the volume integral equation of electromagnetic scattering on a nonmagnetic object of finite spatial extent $D$:
 \begin{align}
 \label{eq:GradDivEFIE}
 \begin{split}
 \bE^{\rm in}(\bx,\omega) = \;& \bE(\bx,\omega) -
 \\ 
 &\left[k_{0}^{2}(\omega) + \nabla\nabla\cdot\right]
 \int\limits_{\bx'\in D}{g(\bx-\bx',\omega)\chi(\bx',\omega)
 \bE(\bx',\omega)\;{\rm d} V},
 \end{split}
 \end{align}
where $\bE^{\rm in}$ and $\bE$ are the incident and total electric fields, correspondingly. 
This equation is obtained directly from the frequency-domain Maxwell's equations and takes into account the 
radiation condition at infinity in a most natural form. 
The medium parameters are contained in the contrast function $\chi$, which in terms of the 
complex permittivity function $\varepsilon$ will look like
 \begin{align}
 \label{eq:Chi}
 \chi(\bx,\omega)=\frac{\varepsilon(\bx,\omega)}{\varepsilon_{0}}-1=\varepsilon_{\rm r}(\bx,\omega)-1.
 \end{align}
The vacuum wavenumber is $k_{0}=\omega/c$, and the scalar Green's function is
given by
 \begin{align}
 \label{eq:ScalarGreen}
 g(\bx,\omega)=\frac{e^{ik_{0}\vert\bx\vert}}{4\pi\vert\bx\vert}.
 \end{align}
Carrying out the two spatial derivatives we 
arrive at the following singular integral equation:
 \begin{align}
 \label{eq:EFVIE}
 \begin{split}
 \bE^{\rm in}(\bx,\omega)
 = 
 &
 \left[1 + \frac{1}{3}\chi(\bx,\omega)\right]
 \bE(\bx,\omega)
 \\
 & - \lim\limits_{\delta\rightarrow 0}
 \int\limits_{\bx'\in D\setminus\vert\bx-\bx'\vert<\delta}
 {\mathbb G}_{0}(\bx-\bx')
 \chi(\bx',\omega)
 \bE(\bx',\omega)
 \;{\rm d}\bx'
 \\
 &- \int\limits_{\bx'\in D}
 {\mathbb G}_{1}(\bx-\bx',\omega)
 \chi(\bx',\omega)
 \bE(\bx',\omega)
 \;{\rm d} \bx' .
 \end{split}
 \end{align}
For the definitions of the Green tensors ${\mathbb G}_{0}$ and ${\mathbb G}_{1}$ we refer to our previous publication 
on this subject \cite{BudkoSamokhinSIAM2006}.

\begin{theorem} 
\label{th:MainTheorem} 
The vector-valued function $\Psi(\alpha,\bx,\bx_{\rm c})$ defined in Theorem~\ref{th:PsiProperties}
generates the essential mode of the electromagnetic field corresponding to the point of essential spectrum
\begin{align}
\label{eq:EssSpec}
\lambda_{\rm ess}=\varepsilon_{\rm r}(\bx_{\rm c},\omega).
\end{align}
\end{theorem} 
\begin{proof}
Since we have already established the normalization (Property~1) and the singularity of 
the sequence (Property~2) in Theorem~\ref{th:PsiProperties}, 
we only need to prove the following analogue of (\ref{eq:DefWeyl}):
 \begin{align}
 \label{eq:SpectralProblem}
 \begin{split}
 V&=\lim\limits_{\alpha\rightarrow\infty} 
 \left\Vert \left[1 + \frac{1}{3}\chi(\bx,\omega)\right]
 \Psi(\alpha,\bx,\bx_{\rm c})
 \right.
 \\
 & - \lim\limits_{\delta\rightarrow 0}
 \int\limits_{\bx'\in D\setminus\vert\bx-\bx'\vert<\delta}
 {\mathbb G}_{0}(\bx-\bx')
 \chi(\bx',\omega)
 \Psi(\alpha,\bx',\bx_{\rm c})
 \;{\rm d}\bx'
 \\
 &\left. - \int\limits_{\bx'\in D}
 {\mathbb G}_{1}(\bx-\bx',\omega)
 \chi(\bx',\omega)
 \Psi(\alpha,\bx',\bx_{\rm c})
 \;{\rm d} \bx' - \lambda\Psi(\alpha,\bx,\bx_{\rm c})
 \right\Vert^{2}_{D}
 \\
 &=\lim\limits_{\alpha\rightarrow\infty} 
 \left\Vert 
 \left[\frac{2}{3}+ \frac{1}{3}\varepsilon_{\rm r}(\bx)-\lambda\right]\Psi(\alpha,\bx,\bx_{\rm c}) 
 \right.
 \\
 &\;\;- \lim\limits_{\delta\rightarrow 0}
 \int\limits_{\bx'\in D\setminus V(\delta)}
 {\mathbb G}_{0}(\bx-\bx')
 \left[\varepsilon_{\rm r}(\bx')-\varepsilon_{\rm r}(\bx)+\varepsilon_{\rm r}(\bx)-1\right]
 \Psi(\alpha,\bx',\bx_{\rm c})
 \;{\rm d}\bx'
 \\
 &\;\;\left.- \int\limits_{\bx'\in D}
 {\mathbb G}_{1}(\bx-\bx',\omega)
 \left[\varepsilon_{\rm r}(\bx')-1\right]
 \Psi(\alpha,\bx',\bx_{\rm c})
 \;{\rm d} \bx'
 \right\Vert_{D}^{2} =  0\,,
 \end{split}
 \end{align}
for $\lambda=\varepsilon_{\rm r}(\bx_{\rm c},\omega)$. The $L_{2}$ norm is taken over the finite 
spatial support $D$. First, we rearrange (\ref{eq:SpectralProblem}) and decompose it into separate terms
 \begin{align}
 \label{eq:SpectralNorm}
 \begin{split}
 V&\le\lim\limits_{\alpha\rightarrow\infty} 
 \left\Vert 
 \left[\varepsilon_{\rm r}(\bx)-\lambda\right]\Psi(\alpha,\bx,\bx_{\rm c}) 
 \right\Vert^{2}_{D}
 \\
 &+\lim\limits_{\alpha\rightarrow\infty} 
 \left\Vert 
 \left[1-\varepsilon_{\rm r}(\bx)\right]
 \left[\frac{2}{3}\Psi(\alpha,\bx,\bx_{\rm c}) 
 +\lim\limits_{\delta\rightarrow 0}
 \int\limits_{\bx'\in D\setminus V(\delta)}
 {\mathbb G}_{0}(\bx-\bx')
 \Psi(\alpha,\bx',\bx_{\rm c})
 \;{\rm d}\bx'
 \right]
 \right\Vert^{2}_{D}
 \\
 &+\lim\limits_{\alpha\rightarrow\infty} 
 \left\Vert 
 \lim\limits_{\delta\rightarrow 0}
 \int\limits_{\bx'\in D\setminus V(\delta)}
 {\mathbb G}_{0}(\bx-\bx')\left[\varepsilon(\bx,\omega)-\varepsilon(\bx',\omega)\right]
 \Psi(\alpha,\bx',\bx_{\rm c})
 \;{\rm d}\bx'
 \right.
 \\
 &\;\;\;\;\;\;\;\;\;\;\;\;\;\;\;\;\;\;\left.- \int\limits_{\bx'\in D}
 {\mathbb G}_{1}(\bx-\bx',\omega)
 \left[\varepsilon_{\rm r}(\bx')-1\right]
 \Psi(\alpha,\bx',\bx_{\rm c})
 \;{\rm d} \bx'
 \right\Vert^{2}_{D}.
 \end{split}
 \end{align}
With $\varepsilon_{\rm r}(\bx,\omega)$ H{\"o}lder-continuous in ${\mathbb R}^{3}$ all integral
operators in the last term are weakly singular. Hence, from the Property~6 of Theorem~\ref{th:PsiProperties}
the last term in (\ref{eq:SpectralNorm}) is zero.

From the second term in (\ref{eq:SpectralNorm}) we only consider the expression in the large square brackets, which we extend to 
${\mathbb R}^{3}$. Denoting by ${\mathcal F}\left\{\dots\right\}$ and ${\mathcal F}^{-1}\left\{\dots\right\}$
the forward and inverse three-dimensional Fourier transforms, we obtain
 \begin{align}
 \label{eq:ExtSingular}
 \begin{split}
 &\frac{2}{3}\Psi(\alpha,\bx,\bx_{\rm c}) 
 +\lim\limits_{\delta\rightarrow 0}
 \int\limits_{\bx'\in {\mathbb R}^{3}\setminus V(\delta)}
 {\mathbb G}_{0}(\bx-\bx')
 \Psi(\alpha,\bx',\bx_{\rm c})
 \;{\rm d}\bx'
 \\
 &={\mathcal F}^{-1}
 \left\{\frac{2}{3}{\mathcal F}\left\{\Psi(\alpha,\bx,\bx_{\rm c})\right\}
 +
 {\mathcal F}\left\{\lim\limits_{\delta\rightarrow 0}
 \int\limits_{\bx'\in {\mathbb R}^{3}\setminus V(\delta)}
 {\mathbb G}_{0}(\bx-\bx')
 \Psi(\alpha,\bx',\bx_{\rm c})
 \;{\rm d}\bx'\right\}
 \right\}
 \\
 &={\mathcal F}^{-1}
 \left\{\frac{2}{3}\tilde{\Psi}(\alpha,\bk,\bx_{\rm c})
 +
 \left[\frac{1}{3}{\mathbb I} - \tilde{\mathbb Q}\right]
 \tilde{\Psi}(\alpha,\bk,\bx_{\rm c})
 \right\}
 \\
 &={\mathcal F}^{-1}
 \left\{\tilde{\Psi}(\alpha,\bk,\bx_{\rm c})
 -\frac{\bk\bk^{\rm T}}{\vert\bk\vert^{2}}
 \tilde{\Psi}(\alpha,\bk,\bx_{\rm c})
 \right\}=0,
 \end{split}
 \end{align}
where we have used the previously derived result about the Fourier transform of a simple singular operator 
\cite{BudkoSamokhinSIAM2006}, the explicit form of $\tilde{\Psi}(\alpha,\bk,\bx_{\rm c})$ (Property~3), and the 
fact that $\bk^{\rm T}\bk=\vert\bk\vert^{2}$. 
Hence, the second term in (\ref{eq:SpectralNorm}) is also zero. 

Finally, applying Property~4 (sifting) of Theorem~\ref{th:PsiProperties} we see that the first term in (\ref{eq:SpectralNorm})
is zero, if equality (\ref{eq:EssSpec}) holds.
\end{proof}

\section{Conclusions}
%%%%%%%%%%%%%%%%%%%%%%%%%%%%%%%%%%%%%%%%%%%%%%%%%%%%%%%%%%%%%%%%%%%%%%%%%%%%%%
%%%%%%%%%%%%%%%%%%%%%%%%%%%%%%%%%%%%%%%%%%%%%%%%%%%%%%%%%%%%%%%%%%%%%%%%%%%%%%
Now we know that the electromagnetic essential (singular) mode is, in fact, the square root of the delta function. 
But what does it mean from the physical point of view? Is it possible to have an ``essential'' resonance? Under which conditions?
Can the electromagnetic field be confined to what seems to be a single point in space? These are open questions. 
However, one important conclusion can be deduced from the basic principles of
the electromagnetic theory.

The resonant excitation of a singular mode is only possible if the essential spectrum contains zero. Physically this
means that $\varepsilon_{\rm r}(\bx_{\rm c},\omega)=0$ at some point $\bx_{\rm c}$ in the scattering domain $D$.
It may seem improbable, but one has to remember that the dielectric permittivity is an effective macroscopic 
parameter, which has very little meaning for the microscopic induced current densities at
the atomic level. Moreover, in the classical Lorentz theory of atoms, the real part of the dielectric permittivity
in a dispersive medium can vanish and even become negative (the so-called Lorentz absorption line).
Although, in that case the Kramers-Kronig relations predict an increase in the imaginary part of macroscopic 
$\varepsilon_{\rm r}$, which corresponds to the absorption of the electromagnetic field. Turning it all around we can say 
that even if at a microscopic level we could have $\varepsilon_{\rm r}(\bx_{\rm c},\omega)=0$ and excite the 
corresponding singular mode, it should be absorbed to preserve the macroscopic Kramers-Kronig relations.

In principle, the very fact that singular modes are outside the Hilbert space, where all `proper' solutions of the Maxwell
equations live, tells us that the excitation of singular modes can be considered as some kind of `deflation' of the 
electromagnetic field. Recall that, due to the normalization Property~1 of Theorem~\ref{th:PsiProperties}, the 
electromagnetic energy associated with a singular mode is well defined.

It is also interesting to note the direct relation of the electromagnetic essential spectrum and its singular modes 
to the pseudospectrum and wave-packet pseudomodes \cite{Trefethen2005}. There is an obvious similarity of the
Weyl's definition (\ref{eq:DefWeyl}) and the definition of the pseudospectrum, where instead of zero one should simply put 
a small $\epsilon$ in the right-hand side of (\ref{eq:DefWeyl}). Subsequently, we arrive at two distinct possibilities.
The first is where $n\rightarrow\infty$, i.e., in our case $\alpha\rightarrow\infty$. Then, points $\lambda_{\rm ps}$ 
satisfying $\vert\lambda_{\rm ps}-\varepsilon_{\rm r}(\bx,\omega)\vert\le \epsilon$ will belong to the 
pseudospectrum, while the corresponding modes will be singular. The second case is 
where $\lambda_{\rm ps}=\lambda_{\rm ess}=\varepsilon_{\rm r}(\bx,\omega)$ or very close to it,
but $\alpha\le \delta(\epsilon)$. In this case, we stop the sequence of $\Psi(\alpha,\bx_{\rm c},\bx)$, at some
finite $\alpha$, for which the norm in (\ref{eq:SpectralProblem}) equals $\epsilon$. Although, it is difficult to derive an explicit 
relation for $\delta(\epsilon)$, we can anticipate that $\Psi(\delta(\epsilon),\bx_{\rm c},\bx)$ will be
highly localized in space around the point $\bx_{\rm c}$. In this case the mode is not singular and belongs to the 
Hilbert space. These two physically distinct possibilities emphasize 
the nonunique nature of the pseudospectrum as it is defined in \cite{Trefethen2005} and elsewhere.


\begin{thebibliography}{10} 
%%%%%%%%%%%%%%%%%%%%%%%%%%%%%%%%%%%%%%%%%%%%%%%%%%%%%%%%%%%%%%%%%%%%%%%%%%%%%%
%%%%%%%%%%%%%%%%%%%%%%%%%%%%%%%%%%%%%%%%%%%%%%%%%%%%%%%%%%%%%%%%%%%%%%%%%%%%%%

\bibitem{Collin} 
{\sc R.~E.~Collin}, 
{\em Foundations for Microwave Engineering}, 
McGraw-Hill Education, 1992.

\bibitem{AmariTriki2003} 
{\sc H.~Amari and F.~Triki}, 
{\it Resonances for microstrip transmission lines'},
SIAM J. Appl. Math., Vol.~64, No.~2, pp.~601--636, 2003.

\bibitem{KartchevskiNosichHanson2005} 
{\sc E.~M.~Kartchevski, A.~I.~Nosich, and G.~W.~Hanson},
{\it Mathematical analysis of the generalized natural modes of an inhomogeneous optical fiber},
SIAM J. Appl. Math., Vol.~65, No.~6, pp.~2033--2048, 2005.

\bibitem{FigotinKuchment1998} 
{\sc A.~Figotin and P.~Kuchment}, 
{\it Spectral properties of classical waves in high-contrast periodic media},
SIAM J. Appl. Math., Vol.~58, No.~2, pp.~683--702, 1998.

\bibitem{ShipmanVenakides2003} 
{\sc S.~P.~Shipman and S.~Venakides}, 
{\it Resonance and bound states in photonic crystal slabs},
SIAM J. Appl. Math., Vol.~64, No.~1, pp.~322--342, 2003.

\bibitem{BudkoSamokhinPRL2006} 
{\sc N.~V.~Budko and A.~B.~Samokhin}, 
{\it Classification of electromagnetic resonances in finite inhomogeneous three-dimensional structures}, 
Phys. Rev. Lett.,  Vol.~96, 023904, 2006.

\bibitem{BudkoSamokhinSIAM2006} 
{\sc N.~V.~Budko and A.~B.~Samokhin}, 
{\it Spectrum of the volume integral operator of electromagnetic scattering}, 
SIAM J. Sci. Comput., Vol.~28, No.~2, pp.~682--700, 2006.

\bibitem{MartinGirardDereux1995} 
{\sc O.~J.~F.~Martin, C.~Girard, and A.~Dereux},
{\it Generalized field propagator for electromagnetic scattering and light confinement},
Phys. Rev. Lett., Vol.~74, 526-–529, 1995.

\bibitem{Maksimenko2000} 
{\sc S.~ A.~Maksimenko, G.~Ya.~Slepyan, N.~N.~Ledentsov, V.~P.~Kalosha, A.~Hoffmann, and D.~Bimberg}, 
{\it Light confinement in a quantum dot},
Semicond. Sci. Technol.,  Vol.~15, 491-–496, 2000.

\bibitem{Akahane2003} 
{\sc Y.~Akahane, T.~Asano, B.-S.~Song, and S.~Noda}, 
{\it High-Q photonic nanocavity in a two-dimensional photonic crystal}, 
Nature, Vol.~425, 944--947, 2003.

\bibitem{Milton2002} 
{\sc G.~W.~Milton}, 
{\em The theory of composites}, 
Cambridge University Press, Cambridge, UK, 2002.

\bibitem{Milton2005} 
{\sc G.~W.~Milton, N.-A.~P.~Nicorovici, R.~C.~McPhedran, V.~A.~Podolskiy},
{\it A proof of superlensing in the quasistatic regime, and limitations of superlenses in this regime 
due to anomalous localized resonance}, 
Proceedings of the Royal Society A: Mathematical, 
Physical and Engineering Sciences, Vol.~461, pp.~3999–-4034, 2005. 

\bibitem{Mikhlin2} 
{\sc S.~G.~Mikhlin and S.~Pr{\"o}ssdorf},
{\em Singular Integral Operators}, 
Springer-Verlag, Berlin, 1986.

\bibitem{HislopSigal1996} 
{\sc P.~D.~Hislop and I.~M.~Sigal},
{\em Introduction to Spectral Theory: With Applications to Schr{\"o}dinger Operators}, 
Springer-Verlag, New York, 1996.

\bibitem{DemuthKrishna2005} 
{\sc M.~Demuth and M.~Krishna},
{\em Determining Spectra in Quantum Theory}, 
Birkh{\"a}user, Boston, 2005.

\bibitem{Byron1992} 
{\sc F.~W.~Byron,~Jr. and R.~W.~Fuller}, 
{\em Mathematics of Classical and Quantum Physics},
Dover, New York, 1992.

\bibitem{Peres1993} 
{\sc A.~Peres}, 
{\em Quantum Theory: Concepts and Methods},
Kluwer Academic Publishers, Dordrecht, 1993.

\bibitem{Colombeau1992} 
{\sc J.~F.~Colombeau}, 
{\em Multiplication of Distributions: A Tool In Mathematics, Numerical Engineering and Theoretical Physics}, 
Springer-Verlag, Berlin, 1992.

\bibitem{Trefethen2005} 
{\sc L.~N.~Trefethen and M.~Embree}, 
{\em Spectra and Pseudospectra: The Behavior of Nonnormal Matrices and Operators}, 
Princeton University Press, 2005.

\end{thebibliography}
\end{document}